\documentclass[preprint]{aastex701}

\newcommand{\dg}{^{\circ}}
\newcommand{\NTOT}{15}
\received{May 29, 2026}
\revised{June 24, 2026}
\accepted{July 6, 2026}

\begin{document}

\title{Spectroscopic redshifts of selected flat-spectrum radio sources I}

\author[0000-0003-0611-5784]{Dmitry Blinov}
\altaffiliation{Corresponding Author: blinov@ia.forth.gr}
\affiliation{Institute of Astrophysics, Foundation for Research and Technology – Hellas, N. Plastira 100, Voutes GR-70013, Heraklion, Greece}
\affiliation{University of Crete, Department of Physics \& Institute of Theoretical \& Computational Physics, 70013 Heraklion, Greece}
\email{blinov@ia.forth.gr}

\author[0009-0004-1671-5454]{Alberto Floris}
\affiliation{University of Crete, Department of Physics \& Institute of Theoretical \& Computational Physics, 70013 Heraklion, Greece}
\affiliation{Institute of Astrophysics, Foundation for Research and Technology – Hellas, N. Plastira 100, Voutes GR-70013, Heraklion, Greece}
\affiliation{National Institute for Astrophysics (INAF), Astronomical Observatory of Padova, IT-35122 Padova, Italy}
\email{afloris@physics.uoc.gr}

\author[0000-0001-8952-676X]{Andreas Zezas}
\affiliation{Institute of Astrophysics, Foundation for Research and Technology – Hellas, N. Plastira 100, Voutes GR-70013, Heraklion, Greece}
\affiliation{University of Crete, Department of Physics \& Institute of Theoretical \& Computational Physics, 70013 Heraklion, Greece}
\email{azezas@ia.forth.gr}

\author[0000-0003-1117-2863]{Carolina Casadio}
\affiliation{Institute of Astrophysics, Foundation for Research and Technology – Hellas, N. Plastira 100, Voutes GR-70013, Heraklion, Greece}
\affiliation{University of Crete, Department of Physics \& Institute of Theoretical \& Computational Physics, 70013 Heraklion, Greece}
\email{ccasadio@ia.forth.gr}

\author[0000-0003-1497-1134]{Elias Kyritsis}
\affiliation{University of Crete, Department of Physics \& Institute of Theoretical \& Computational Physics, 70013 Heraklion, Greece}
\affiliation{Institute of Astrophysics, Foundation for Research and Technology – Hellas, N. Plastira 100, Voutes GR-70013, Heraklion, Greece}
\email{ekyritsis@physics.uoc.gr}

\author[0000-0001-8382-3229]{Valentina Missaglia}
\affiliation{INAF – Osservatorio di Astrofisica e Scienza dello Spazio di Bologna, Via Gobetti 101, I-40129 Bologna, Italy}
\email{valentina.missaglia@inaf.it}

\author[]{Vangelis Pantoulas}
\affiliation{Institute of Astrophysics, Foundation for Research and Technology – Hellas, N. Plastira 100, Voutes GR-70013, Heraklion, Greece}
\email{vpan@ia.forth.gr}

\author[0000-0002-7994-5167]{Avinash Kumar}
\affiliation{Institute of Astrophysics, Foundation for Research and Technology – Hellas, N. Plastira 100, Voutes GR-70013, Heraklion, Greece}
\affiliation{University of Crete, Department of Physics \& Institute of Theoretical \& Computational Physics, 70013 Heraklion, Greece}
\email{paleolo@physics.uoc.gr}


\begin{abstract}

We present the first results of a spectroscopic campaign carried out as part of the Search for Milli-Lenses (SMILE) program, which aims to constrain the prevalence of gravitational lens systems on milli-arcsecond angular scales (milli-lenses) using high-resolution Very Long Baseline Interferometry (VLBI) imaging. The SMILE parent sample contains $\sim 5000$ radio-loud active galaxies, selected as a flux-limited, complete subsample of CLASS (The Cosmic Lens All-Sky Survey) sources. We compiled redshift information for the full sample from multiple literature and catalog sources and found that 491 sources have no available redshift estimate, either spectroscopic or photometric. A further 948 sources have only photometric redshifts, many of
which show substantial discrepancies between catalogs. Reliable redshifts are essential for VLBI radio-source studies because they convert angular measurements into physical linear scales, enable estimates of intrinsic luminosities and jet kinematics, and allow robust cosmological and population studies. To address this key limitation for lensing and population studies, we initiated a dedicated spectroscopic campaign to secure reliable redshifts for as many targets as possible. This paper focuses on the brightest sources in the SMILE sample. We report newly determined spectroscopic redshifts for 6 targets out of \NTOT{} observed with the Skinakas 1.3 m telescope.

\end{abstract}



\section{Introduction} \label{sec:Intro}

The Search for Milli-Lenses (SMILE) project\footnote{\url{https://smilescience.info/}} \citep{Casadio2021} is designed to search for gravitational lens systems on milli-arcsecond scales (milli-lenses) using high-resolution Very Long Baseline Interferometry (VLBI) data for a large, complete sample of radio-loud active galaxies (AGN). Milli-arcsecond image separations probe lensing by supermassive compact objects with masses in the range $10^6$--$10^9\,M_\odot$. This mass regime is especially important in the context of the widely accepted $\Lambda$ cold dark matter ($\Lambda$CDM) model, which predicts substantially more dark matter halos at subgalactic scales ($\le 10^{11}\,M_\odot$) than are currently observed (see \cite{Zavala&Frenk2019} for a review).

For SMILE, we created a flux-limited sample (flux density at 8 GHz $\ge 50$ mJy) of 4968 radio sources, starting from the complete sample of 11685 radio-loud sources in the Cosmic Lens All-Sky Survey \citep[CLASS,][]{Myers2003,Browne2003}. The CLASS catalog is drawn from two other catalogs (5 GHz GB6 catalog \citep{Gregory1996} and NVSS 1.4 GHz catalog \citep{Condon1998}) and contains sources in the declination range $[0\dg, +75\dg]$, with flux density $\ge 30$ mJy at 5 GHz, flat spectral index ($\alpha <0.5$ with $S_{\nu}\propto\nu^\alpha$) between 1.4 and 5 GHz, and Galactic latitude (b) $>$ 10$\dg$.

Reliable redshifts are essential for several parts of the SMILE analysis. Lensing cross-sections used in statistical studies depend on source distances \citep{Loudas2022}. A redshift is required to convert observed angular structure into linear size, which is critical for rejecting false milli-lens candidates. One major contaminant in milli-lens searches is the compact symmetric object (CSO) population \citep{Potzl2025}, whose members are typically characterized by projected sizes $<1$ kpc \citep{Kiehlmann2024}. Therefore, it is crucial to determine the linear separation between compact radio components in order to distinguish possible milli-lenses from CSOs. Moreover, redshifts are needed to derive the intrinsic luminosities of radio components and to model any confirmed milli-lens systems. For these reasons, redshifts with at least photometric accuracy are highly desirable across the SMILE sample, though spectroscopic redshifts ($z_{\rm spec}$) are strongly preferred wherever achievable, as their higher precision reduces systematic uncertainties in size estimates and lens modeling. Maximizing the number of sources with secure $z_{\rm spec}$ therefore remains an important goal of SMILE.

We therefore carried out an extensive redshift compilation using major surveys and catalogs. We cross-matched the SMILE sample with the Optical Characteristics of Astrometric Radio Sources catalog \citep[OCARS,][]{Malkin2016}, the Sloan Digital Sky Survey Data Release 18 \citep[SDSS DR18,][]{Almeida2023}, the Million Quasars Catalogue \citep[Milliquas v8,][]{Flesch2023}, the Pan-STARRS1 Source Types and Redshifts with Machine Learning catalog \citep[PS1-STRM,][]{Beck2021}, the Dark Energy Spectroscopic Instrument (DESI) Legacy Imaging Surveys DR8 and DR10 \citep{Dey2019}, the DESI Spectroscopic Data Release 1 \citep[DESI DR1,][]{Karim2025}, the Large Sky Area Multi-Object Fiber Spectroscopic Telescope (LAMOST) Quasar Survey DR1--DR12 \citep{Ai2016,Dong2018,Yao2019,Jin2023,Lyu2026}, the Catalog of 5 Million Quasars from the Zwicky Transient Facility \citep[QZO,][]{Nakoneczny2025}, and the CatGlobe catalog \citep{Fu2025}, with additional checks in SIMBAD \citep{Wenger2000A} and the NASA/IPAC Extragalactic Database \citep[NED,][]{Helou1991}. Even after this effort, 948 sources have only photometric redshifts and 491 have no redshift information at all. Moreover, even in the era of big data and machine learning, when individual projects \citep[e.g., Gaia;][]{BailerJones2023} are capable of providing redshift measurements for millions of extragalactic objects, manual inspection of individual observations still yields results of unparalleled accuracy. In practice, the aforementioned catalogs often report inconsistent redshifts for the same object, as automated algorithms can suffer from catastrophic failures, even in spectroscopic surveys.

To address this gap, we initiated a dedicated spectroscopic follow-up program using multiple optical facilities. The median G-band (Gaia DR3) magnitude of SMILE sources with known $z_{\rm spec}$ is 19.3, whereas sources lacking spectroscopic redshifts (either having only photometric estimates or no estimate at all) are typically fainter, with a median of 20.5. Nevertheless, the absence of $z_{\rm spec}$ is not solely a matter of brightness: a significant fraction of these sources were never observed spectroscopically, or were observed during a state in which emission lines were too weak to be detected. In fact, more than 100 sources with $G<20$ remain without $z_{\rm spec}$, making them accessible to small-aperture telescopes. Our follow-up program specifically targets this population. In this paper, we present the first results from observations with the 1.3 m Skinakas telescope: new spectroscopic redshift determinations for 6 AGN.

\section{Observations and data reduction} \label{sec:Obs}

From the SMILE sample, we selected sources without spectroscopic redshift measurements with $r<19^m$ according to the Pan-STARRS (PS1) DR1 catalog \citep{Chambers2016} or $G_{\rm RP}<19^m$ according to Gaia DR2 \citep{Riello2018}, depending on the availability of photometric measurements. At the time of each observing run, the brightest available source from this subsample was selected for observation. We note that for jet-dominated AGN, the optical continuum is often governed by highly variable synchrotron emission, which can partially or completely obscure the emission lines required for redshift determination. Consequently, observations obtained during a bright optical state are not necessarily more favorable for spectroscopic redshift measurements.

The observations were carried out in 2022--2023 with the 1.3 m telescope at Skinakas Observatory (Crete, Greece). The instrumental setup included an ANDOR iKon-L 936 BEX2-DD back-illuminated CCD (2048 $\times$ 2048 pixels; 13.5 $\mu$m pixel size) and a spectrograph equipped with a 600 lines mm$^{-1}$ grating, providing a nominal dispersion of $\sim 2$ \AA\ pixel$^{-1}$. The spectra were obtained with a 160\,$\mu$m-wide slit centered at each object. Exposure times typically ranged between 1000 and 5500\,s, depending on the brightness of the target. Because absolute flux measurements were not required, photometric conditions were not necessary. The typical spectral resolving power of our observations was $R = \lambda / \Delta \lambda \approx 530$.

We reduced the spectra with a custom Python pipeline that leverages the \texttt{astropy}, \texttt{ccdproc}, \texttt{astroscrappy}, \texttt{scipy}, and \texttt{numpy} packages. All frames were bias-subtracted and flat-field corrected. The master flat-field was produced by bias-subtracting tungsten-lamp exposures, sigma-clipping the stack at the $3\sigma$ level, and median-combining the result. The combined flat was then normalized by collapsing it along the spatial axis to a one-dimensional profile, fitting a fourth-degree polynomial to remove the lamp's spectral energy distribution, and dividing the two-dimensional flat by the resulting smooth function, leaving only the pixel-to-pixel response. Cosmic rays in each individual science frame were identified and replaced using the L.A.Cosmic algorithm \citep[astroscrappy,][]{vanDokkum2001}. Because the spectral trace was found to be tilted on the detector by a small angle (measured interactively by fitting a straight line to the continuum centroid at several positions along the dispersion axis), all combined frames were rotated to align the dispersion axis with the detector rows before further processing. Comparison-lamp (Photron P826) exposures were obtained before and after each science exposure to account for possible small wavelength-calibration drifts during the night. Wavelength calibration was performed by extracting a one-dimensional arc spectrum along the trace, automatically detecting emission-line peaks, and cross-matching them against a reference list of $\sim15$ He/Ne/Ar lines spanning approximately 5850–8520 \AA. A second-degree polynomial was fitted to the matched pixel–wavelength pairs; the root-mean-square residual of this solution was recorded for each exposure. One-dimensional spectra were extracted using an aperture centered on the Gaussian-fitted trace centroid, enclosing more than 90\% of the source flux. Sky emission was subtracted using two source-free background windows located around the aperture edges after a gap of a few pixels; the median of each window was scaled to the aperture width and subtracted column by column. Flux calibration was not performed, because our analysis requires only the identification and positions of spectral lines instead of flux measurements. The spectra were normalized to the local continuum by fitting and subtracting a fourth-degree polynomial. This fit was performed iteratively (6 iterations, $2.6\sigma$ clipping threshold), with emission and absorption features progressively excluded at each iteration through sigma clipping.

To measure the line positions, we modeled each profile with a Gaussian function. For blended lines, we fitted up to three Gaussian components.
The redshift was determined as the weighted mean of the redshifts measured from individual lines. The uncertainty on each individual line redshift was derived by propagating the uncertainties from both the dispersion solution fit and the line centroid measurement. The redshift uncertainty was then estimated as the standard error of the mean of these individual line redshifts.

\section{Results} \label{sec:res}

In this section, we summarize the spectroscopic redshift measurements for the six sources with secure line identifications.

\subsection{J010341+423925}

\begin{figure}[ht!]
\plotone{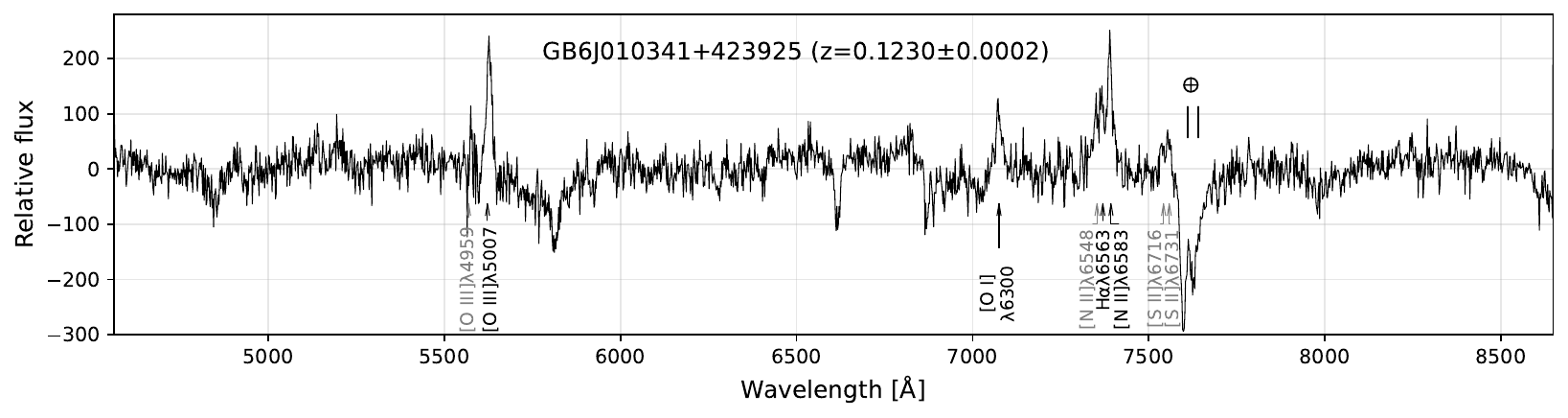}
\caption{Spectrum of GB6 J010341+423925 with identified lines.
\label{fig:J0103}}
\end{figure}

GB6 J010341+423925 (RA = 01:03:40.1, Dec = 42:39:35.6) is positionally consistent with a source of $r=17.92^m$ in the Pan-STARRS Data Release 2 \citep[hereafter PS1-DR2,][]{Magnier2020}. Photometric redshift estimates are $z=0.096\pm0.005$ (PS1-STRM), $z=0.113387$ in the 2MASS Photometric Redshift catalog \citep[2MPZ,][]{Bilicki2014} and $z=0.135\pm0.007$ (SDSS DR18). Our spectrum, obtained with a 5500 s exposure, is shown in Figure~\ref{fig:J0103}. The signal-to-noise ratio (SNR) per pixel ranges from 3 to 18, with an average value of 11.
We identified the [O~III] $\lambda5007\mathrm{\AA}$, [O~I] $\lambda6300\mathrm{\AA}$, H$_\alpha$ and N~II $\lambda 6583\mathrm{\AA}$ lines and derived a redshift of $z=0.1230\pm0.0002$ based on these lines. A few other less prominent lines are visible in the spectrum; these are consistent with [O~III] $\lambda4959\mathrm{\AA}$ and the S~II ($\lambda\lambda 6716,6731\mathrm{\AA}$) blend based on the measured z. Our spectroscopic redshift is comfortably close to the mean of the photometric estimates.

\subsection{J184835+213156}

\begin{figure}[ht!]
\plotone{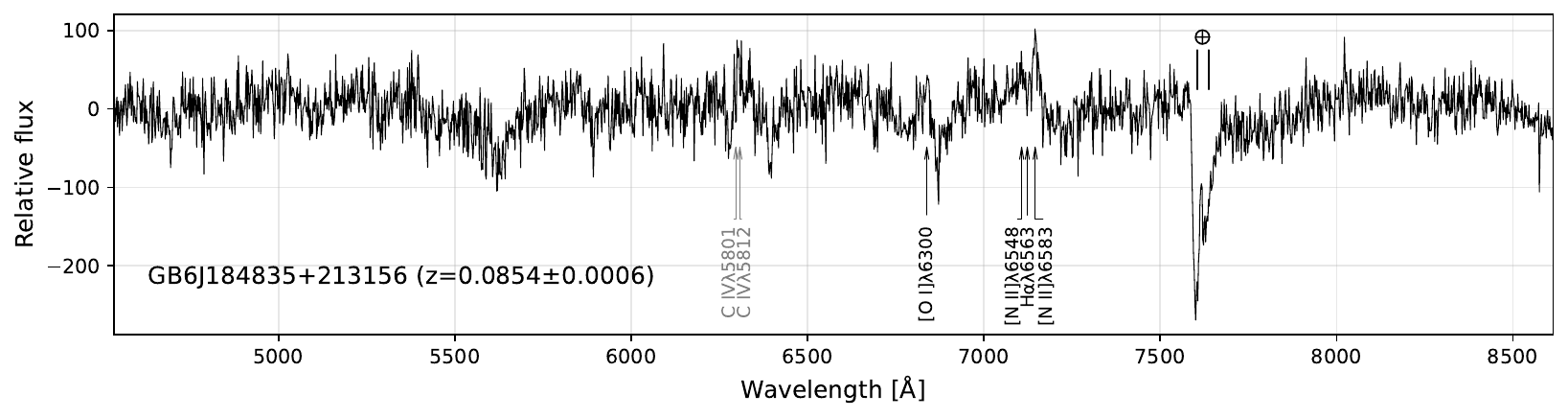}
\caption{Spectrum of GB6 J184835+213156 with identified lines.
\label{fig:J1848}}
\end{figure}

GB6 J184835+213156 (RA = 18:48:36.5, Dec = 21:31:58.5) is positionally consistent with a source of $r=17.89^m$ in PS1-DR2. Two photometric redshift estimates are available: $z=0.069\pm0.002$ (PS1-STRM) and $z=0.0802$ (2MPZ). Our spectrum, obtained with a 2700 s exposure, is shown in Figure~\ref{fig:J1848}. The SNR per pixel ranges from 3 to 19, with an average value of 12.
We identified the [O~I] $\lambda6300\mathrm{\AA}$, N~II ($\lambda\lambda 6548,6583\mathrm{\AA}$) doublet and H$_\alpha$. Fitting these lines yielded $z=0.0854\pm0.0006$. Possible C~IV features at $\lambda 5801\mathrm{\AA}$ and $\lambda 5812\mathrm{\AA}$ may also be present, but were not used in the redshift determination. Our spectroscopic redshift value is roughly consistent with the 2MPZ photometric estimate.

\subsection{J195141+480145}

\begin{figure}[ht!]
\plotone{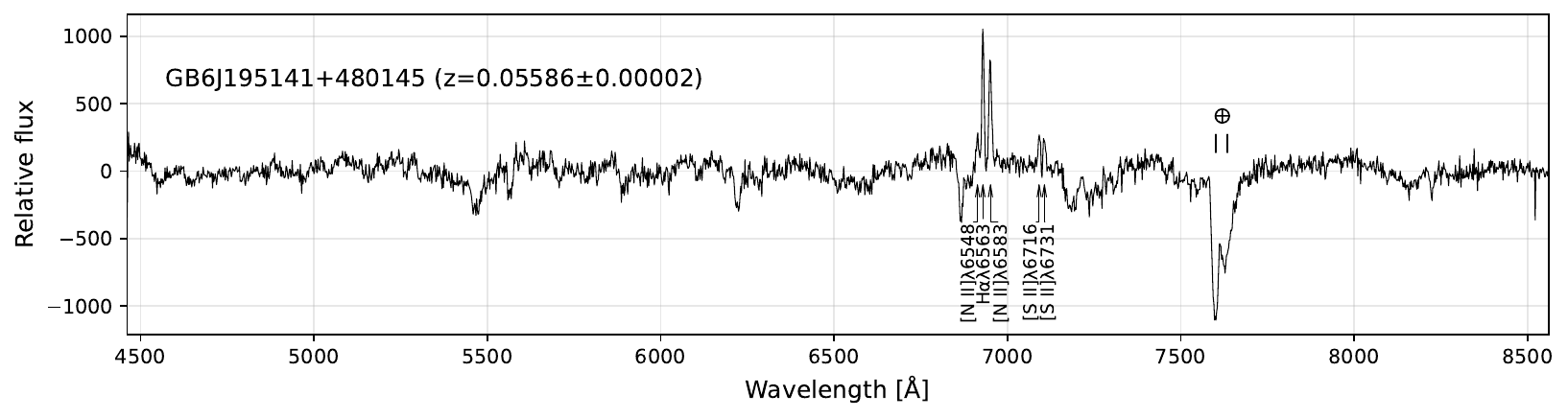}
\caption{Spectrum of GB6 J195141+480145 with identified lines.
\label{fig:J1951}}
\end{figure}

GB6 J195141+480145 (RA = 19:51:41.6, Dec = 48:01:41.4) is positionally consistent with a source of $r=17.36^m$ in PS1-DR2. A photometric redshift of $z=0.021\pm0.003$ is reported in PS1-STRM. Our spectrum, obtained with a 5000 s exposure, is shown in Figure~\ref{fig:J1951}. The SNR per pixel ranges from 4 to 12, with an average value of 9.
We identified the N~II ($\lambda\lambda 6548,6583\mathrm{\AA}$) and S~II ($\lambda\lambda 6716,6731\mathrm{\AA}$) doublets, along with H$_\alpha$. Fitting these lines yields $z=0.05586\pm0.00002$. This value is consistent with the photometric redshift from PS1-STRM given the statistical uncertainties of this catalog (e.g., $\sigma (\Delta z_{\rm norm}) \approx 0.03$).

\subsection{J201414+063439}

\begin{figure}[ht!]
\plotone{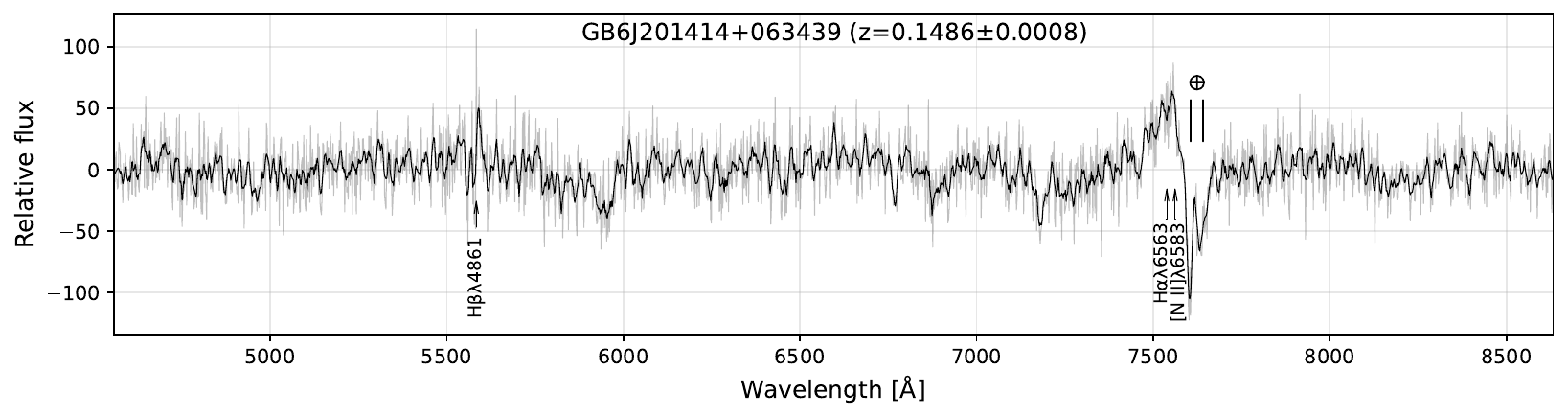}
\caption{Spectrum of GB6 J201414+063439 with identified lines. The original spectrum is shown in gray, while the black line shows the smoothed version.
\label{fig:J2014}}
\end{figure}

GB6 J201414+063439 (RA = 20:14:14.9, Dec = 06:34:37.2) is positionally consistent with a source of $r=18.54^m$ in PS1-DR2. Photometric redshift estimates are $z=0.126\pm0.006$ (PS1-STRM) and $z=0.217$ (CatGlobe). Our spectrum, obtained with a 2490 s exposure, is shown in Figure~\ref{fig:J2014}. For visualization purposes only, we also show a version smoothed using the Savitzky--Golay filter \citep{Savitzky1964}. The SNR per pixel ranges from 2 to 11, with an average value of 6.
Despite the low SNR, we identified H$_\beta$ as well as a broad emission feature that presumably includes H$_\alpha$ and N~II $\lambda 6583\mathrm{\AA}$. Using these lines, we measured $z=0.1486\pm0.0008$, which falls between the two photometric estimates.

\subsection{J203142+162147}

\begin{figure}[ht!]
\plotone{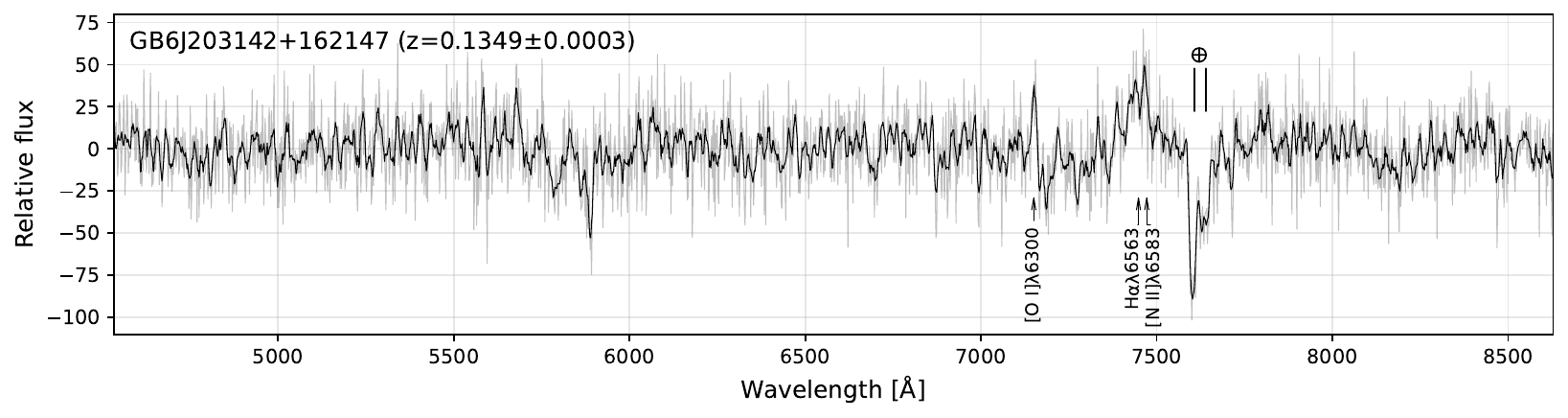}
\caption{Spectrum of GB6 J203142+162147 with identified lines. The original spectrum is shown in gray, while the black line shows the smoothed version.
\label{fig:J2031}}
\end{figure}

GB6 J203142+162147 (RA = 20:31:42.1, Dec = 16:22:07.4) has $r=18.41^m$ in PS1-DR2. Photometric redshift estimates of $0.117\pm0.009$, $0.2687$, and $0.3966$ are reported in PS1-STRM, QZO, and CatGlobe, respectively. Our spectrum obtained with a 2400 s exposure is shown in Figure~\ref{fig:J2031}. The SNR per pixel ranges from 2 to 10, with an average value of 6.
We identified the H$_\alpha$, N~II $\lambda 6583\mathrm{\AA}$, and O~I $\lambda 6300\mathrm{\AA}$. Using these lines, we measured $z=0.1349\pm0.0003$, which lies between the photometric redshifts reported in PS1-STRM and QZO.

\subsection{J222252+144119}

\begin{figure}[ht!]
\plotone{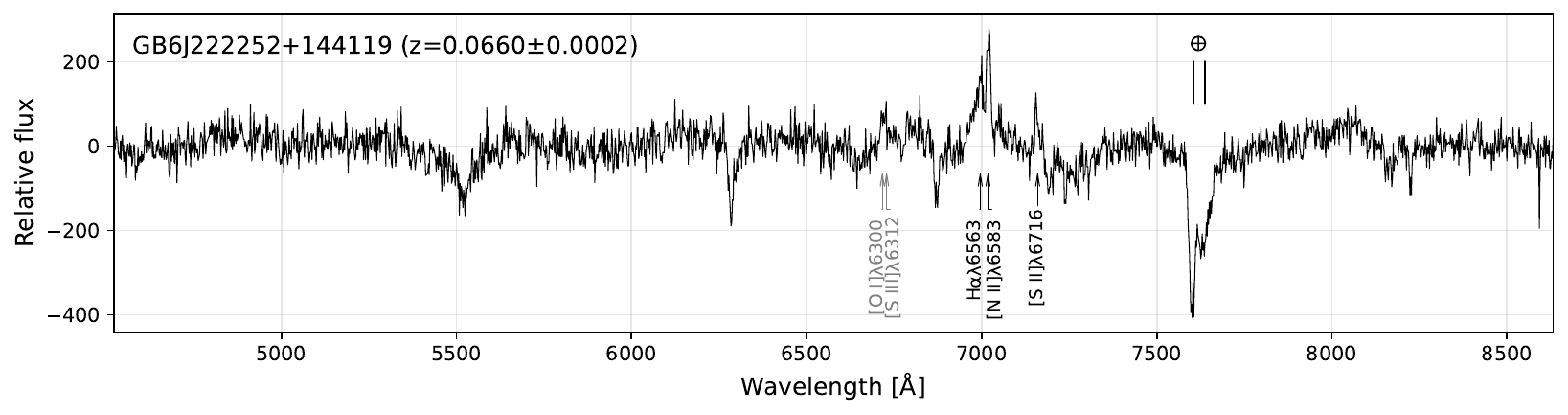}
\caption{Spectrum of GB6 J222252+144119 with identified lines.
\label{fig:J2222}}
\end{figure}

GB6 J222252+144119 (RA = 22:22:51.4, Dec = 14:41:13.6) is positionally consistent with a prominent galaxy of $r=17.02^m$ in PS1-DR2. Published photometric redshifts include $z=0.056\pm0.011$ (SDSS DR18), $z=0.0702\pm0.004$ (PS1-STRM), $z=0.064\pm0.022$ (DESI LS8), $z=0.0749\pm0.0089$ (DESI LS10), and $z=0.054931$ (2MPZ). Our spectrum, obtained with a 1300 s exposure, is shown in Figure~\ref{fig:J2222}. The SNR per pixel ranges from 7 to 26, with an average value of 17.
Using H$_\alpha$, [N~II] $\lambda6583\mathrm{\AA}$, and [S~II] $\lambda6716\mathrm{\AA}$, we measured $z=0.0660\pm0.0002$. A possible [O~I] $\lambda6300\mathrm{\AA}$ and [S~III] $\lambda6312\mathrm{\AA}$ blend is also visible, but was not used in the redshift determination. Our redshift estimate is broadly consistent with all photometric estimates.

\subsection{Other sources}

For four additional sources, we obtained optical spectra at a time when no spectroscopic redshifts were available in the literature; such measurements appeared later. We report our measurements here because they provide an independent verification of both our analysis and external catalog data. For GB6 J000114+235801, the DESI Spectroscopic Data Release 1 \citep[DESI DR1,][]{Karim2025} reports $z=0.10361\pm 0.00001$, while we found $z=0.1034\pm0.0004$ by fitting the identified H$_\alpha$, N~II $\lambda 6583\mathrm{\AA}$, and S~II ($\lambda\lambda 6716,6731\mathrm{\AA}$) lines. For GB6 J014335+263917, DESI DR1 reports $z=1.70771\pm0.00005$, while we found $z=1.702\pm0.004$ based on Si III] $\lambda 1892\mathrm{\AA}$, C III] $\lambda 1909\mathrm{\AA}$, and Mg II] ($\lambda\lambda 2796,2803\mathrm{\AA}$). For GB6 J034257+274858, OCARS reports a spectroscopic redshift of $z=1.9574$, citing Quaia \citep{StoreyFisher2024} as the source of this value. We found $z=1.9529\pm0.0008$ based on C III] $\lambda 1909\mathrm{\AA}$, [Mg V] $\lambda 2783\mathrm{\AA}$, and Mg II] ($\lambda\lambda 2796,2803\mathrm{\AA}$). For GB6 J212043+202610, the LAMOST Quasar Survey \citep{Jin2023} provides $z=2.54089$. This source has multiple prominent emission lines, among which we identified O I $\lambda 1302\mathrm{\AA}$, Si IV ($\lambda\lambda 1394,1403\mathrm{\AA}$), Si III] $\lambda 1892\mathrm{\AA}$, and C III] $\lambda 1909\mathrm{\AA}$. Using these lines, we determined $z=2.538\pm0.003$ for this source.

For four other sources, GB6 J193000+591737, GB6 J224753+441316, GB6 J185854+570831, and GB6 J192935+614638, we were unable to determine redshifts because no emission lines could be identified, either due to the intrinsic absence of emission lines (some of these sources may be featureless BL Lacertae-type AGN) or because of insufficient SNR.

For GB6 J182911+272924, the closest optical source ($r=16.7^m$) is located 0.46 arcsec from the radio position. This source was observed by \citet{Shaw2013}, although no redshift was reported. We obtained its spectrum but found no emission lines; the spectrum is consistent with a stellar spectrum with absorption features. In Gaia DR3 \citep{GaiaDR3}, this source shows significant proper motion and parallax. Moreover, its Renormalized Unit Weight Error (RUWE), a parameter that quantifies the quality of the Gaia astrometric solution, is relatively high, possibly indicating the presence of an AGN blended with the star.

\section{Conclusions} \label{sec:conclusions}

We presented the first results of our dedicated spectroscopic follow-up program for the SMILE sample, targeting radio-loud AGN with flat radio spectra that lacked reliable spectroscopic distance measurements. Using observations from the 1.3 m Skinakas telescope, we obtained optical spectra and measured new spectroscopic redshifts for 6 sources out of \NTOT{} observed in this campaign. These new values, together with spectroscopic redshifts that appeared in the literature after our observations, are summarized in Table~\ref{tab:results}, which also includes sources for which no emission lines could be identified.
\begin{deluxetable*}{lcccccccc}
\tablewidth{0pt}
\tablecaption{Spectroscopic measurement results \label{tab:results}}
\tablehead{
\colhead{Source} & \colhead{RA (J2000)} & \colhead{Dec (J2000)} & exposure & r-mag &\colhead{z} & \colhead{Other $z_{\rm sp}$} & \colhead{\begin{tabular}{@{}c@{}} Other $z_{\rm sp}$\\[-2pt] reference\end{tabular}}   & \colhead{Class} \\
                 & \colhead{hh:mm:ss}   & \colhead{dd:mm:ss}    & sec  &       &      &                   &  & 
}
\startdata
\multicolumn{8}{c}{New redshift measurements} \\
J010341+423925 & 01:03:40.1 & 42:39:36 & 5500 & 17.9 & $0.1230\pm0.0002$   & - & - & AL\\
J184835+213156 & 18:48:36.5 & 21:31:59 & 2700 & 17.9 & $0.0854\pm0.0006$   & - & - & G\\
J195141+480145 & 19:51:41.6 & 48:01:41 & 5000 & 17.4 & $0.05586\pm0.00002$ & - & - & G\\
J201414+063439 & 20:14:14.9 & 06:34:37 & 2490 & 18.5 & $0.1486\pm0.0008$   & - & - & AL\\
J203142+162147 & 20:31:42.1 & 16:22:07 & 2400 & 18.4 & $0.1349\pm0.0003$   & - & - & AL\\
J222252+144119 & 22:22:51.4 & 14:41:14 & 1300 & 17.0 & $0.0660\pm0.0002$   & - & - & G\\
\hline
\multicolumn{8}{c}{Confirmed redshift measurements} \\ 
J000114+235801 & 00:01:14.9 & 23:58:11 & 1600 & 17.9 & $0.1034\pm0.0004$   & $0.10361\pm 0.00001$ & 1 & G\\
J014335+263917 & 01:43:37.1 & 26:39:33 & 1800 & 17.5 & $1.702\pm0.004$     & $1.70771\pm0.00005$  & 1 & AQ\\
J034257+274858 & 03:42:58.9 & 27:49:17 & 6400 & 17.2 & $1.9529\pm0.0008$   & $1.9574$             & 2 & G\\
J212043+202610 & 21:20:42.8 & 20:26:27 & 3600 & 17.3 & $2.538\pm0.003$     & $2.54089$            & 3 & AL\\
\hline
\multicolumn{8}{c}{Sources without identified emission lines} \\
J182911+272924 & 18:29:14.0 & 27:29:03 & 5400 & 16.7 & - & - & - & AL$^*$\\
J185854+570831 & 18:58:53.5 & 57:08:10 & 3600 & 18.0 & - & - & - & AL\\
J192935+614638 & 19:29:35.1 & 61:46:29 & 4500 & 17.9 & - & - & - & AL\\
J193000+591737 & 19:30:00.4 & 59:17:33 & 2400 & 18.0 & - & - & - & V\\
J224753+441316 & 22:47:53.2 & 44:13:15 & 5000 & 17.6 & - & - & - & AL\\
\enddata
\tablecomments{Coordinates are given according to CLASS \citep{Myers2003}. External spectroscopic redshift ($z_{\rm sp}$) references: (1) \citep[DESI DR1;][]{Karim2025} (2) OCARS \citep{Malkin2016}; (3) LAMOST Quasar Survey \citep{Jin2023}. Magnitudes are taken from the Pan-STARRS (PS1) DR2 catalog \citep{Magnier2020}. Source classifications follow OCARS, where "G" denotes a radio galaxy, "AL" a BL Lac object, "AQ" a quasar, and "V" a visual source. An asterisk (*) denotes a source whose optical counterpart is identified as a star based on our spectrum.}
\end{deluxetable*}

For each source, we compared our measurements with previously available photometric estimates. Although the overall agreement is good, the spectroscopic values provide significantly more robust distance constraints. In several cases, photometric redshift estimates differ substantially across catalogs, highlighting the need for direct spectroscopy. We further validated our methodology and results using 4 sources for which spectroscopic redshifts became available after our sample selection and observations; our measurements show very good agreement with the external values.

In future papers, we will present additional AGN redshifts derived from spectra obtained with larger telescopes.

\section{Data availability} \label{sec:data}

The normalized spectra presented in this paper will be available in the Harvard Dataverse at \url{https://doi.org/10.7910/DVN/TWHOIR}.

\begin{acknowledgments}
We thank P. Reig for useful comments on data processing. C.C., D.B., V.M. \& A.K. acknowledge support from the European Research Council (ERC) under the Horizon ERC Grants 2021 program under grant agreement No. 101040021. A.F. was funded by the European Union ERC-2022-STG - BOOTES - 101076343. V.M. acknowledges financial support from the Bando Ricerca Fondamentale INAF 2023, Large Program 1.05.23.01.06 ("The XRISM-to-XIFU (X2X) Agreement and Beyond: entering a new Era of High Resolution X-Ray Spectroscopy"). The Photometric Redshifts for the Legacy Surveys (PRLS) catalog used in this paper was produced with funding from the U.S. Department of Energy Office of Science, Office of High Energy Physics, via grant DE-SC0007914.
\end{acknowledgments}

\begin{contribution}

D.B., A.F., E.K., V.M., E.P., and A.K. performed the observations. A.F. and D.B. developed the pipeline for spectral processing and analysis. V.P. developed the software controlling the spectrograph. C.C., as the SMILE PI, defined the scientific goals and selected the sample sources. V.M., D.B. and C.C. cross-matched the sample with various catalogs. D.B., C.C., and A.Z. contributed to the writing of the manuscript. All authors contributed to the discussion and interpretation of the results.


\end{contribution}

%
\facilities{Skinakas:1.3m}

\software{astropy \citep{Robitaille2013,Price2018,Price2022}, ccdproc \citep{Craig2017}, SciPy \citep{Virtanen2020}
          }



\bibliography{Skinspectra}{}
\bibliographystyle{aasjournalv7}



\end{document}